\documentclass[runningheads]{llncs}

\usepackage[T1]{fontenc}
\usepackage{graphicx,subfigure}
\usepackage{amsmath,amssymb}

\begin{document}
\title{Hyperbolic decomposition of Dirichlet distance for ARMA models}
\author{Jaehyung Choi}
\institute{\email{jj.jaehyung.choi@gmail.com}}

\maketitle

\begin{abstract}
	We investigate the hyperbolic decomposition of the Dirichlet norm and distance between autoregressive moving average (ARMA) models. With the K\"ahler information geometry of linear systems in Hardy spaces and weighted Hardy spaces, we demonstrate that the Dirichlet norm and distance of ARMA models, corresponding to the mutual information between the past and future, are decomposed into functions of the hyperbolic distances between the poles and zeros of the ARMA models. Moreover, the distance is also expressed with separate terms from AR parts, MA parts, and AR-MA cross terms. Furthermore, the hyperbolic decomposition is helpful for the model order reduction of ARMA models.
	
\end{abstract}

\section{Introduction}
	Information geometry, an interdisciplinary field that combines information theory, probability theory, statistics, and differential geometry, has gained prominence in various research domains. Notably, time series analysis and signal processing have become significant areas of applications for information geometry, offering both deeper mathematical insights and practical benefits. As the differential-geometric approach to linear systems and time series models was developed \cite{amari1987differential,ravishanker1990differential,ravishanker2001differential}, Bayesian predictive priors were also derived from this geometric framework, demonstrating that these priors outperform Jeffreys' prior \cite{tanaka2018superharmonic}. Furthermore, the priors have been identified for various signal processing filters and time series models \cite{choi2015kahlerian,choi2015application,choi2015geometric,oda2021shrinkage}. 
	
	In information geometry, such mathematical sophistication is equipped with both theoretical and practical advantages. A notable example is the K\"ahler geometry-based approach to the information geometry of time series models and linear systems \cite{barbaresco2006information,choi2015kahlerian}. The mathematical formulation using the K\"ahler manifold for the information geometry of signal filters facilitates efficient computation of geometric objects from the K\"ahler potential \cite{choi2015kahlerian}. Furthermore, the K\"ahler structure within information geometry also provides computational benefits in identifying high-performing Bayesian predictive priors for linear systems \cite{choi2015kahlerian,choi2015application,choi2015geometric,oda2021shrinkage}.
	
	It is important to note that the K\"ahler potential in the K\"ahler information geometry of linear systems corresponds to the squared Hardy norm of a logarithmic transfer function \cite{choi2015kahlerian}. Beyond the Hardy norm of a logarithmic transfer function, the K\"ahler structure of the linear system geometry can be extended to more general function spaces such as weighted Hardy spaces \cite{choi2021k}. Specifically, the weighted Hardy norm of linear systems in Dirichlet space, also known as the mutual information between the past and future of signal filters \cite{martin2000metric}, is expressed with the sum of logarithmic functions of the poles and zeros in the transfer function of the linear system \cite{choi2021k}.
	
	In this paper, we explore the geometric properties of the Dirichlet distance between autoregressive moving average (ARMA) models. We show that the weighted Hardy norm and distance of linear systems in Dirichlet space given in Choi \cite{choi2021k} are represented with functions of hyperbolic distances between the poles and zeros of the ARMA processes. In particular, the Dirichlet norm and distance are decomposed into the sum of the hyperbolic distances corresponding to the autoregressive components, moving average components, and their cross terms between the poles and zeros of the ARMA models. We also discuss its application to model order reduction using Dirichlet distance.
	 
	The structure of this article is as follows. In the next section, we briefly introduce K\"ahler information geometry of linear systems in weighted Hardy spaces including linear systems in Dirichlet space. In section \ref{geometric_interpretation}, we find the hyperbolic decomposition of Dirichlet distances between ARMA processes. After then we conclude the paper.
	
\section{K\"ahler geometry of linear systems in weighted Hardy spaces}
	In this section, we re-visit the information geometry of linear systems in Hardy space and its K\"ahlerization \cite{choi2015kahlerian}. Moreover, its generalization from linear systems in Hardy spaces to linear systems in weighted Hardy spaces \cite{choi2021k} is also presented. We follow Choi and Mullhaupt \cite{choi2015kahlerian} and Choi \cite{choi2021k} along this section.
	
	The transfer function of a linear system encodes critical information of the system on how an input signal is transformed to an output signal. Using $Z$-transformation, the transfer function $h$ of the signal filter is represented with an input signal $X$ and an output signal $Y$:
	\begin{align}
		Y(z) = h(z;\boldsymbol{\xi})X(z),
	\end{align}
	where $z$ is a complex number on the unit disk, and $\boldsymbol{\xi}$ are the parameters of the linear system. 
	
	The characteristic features of each signal filter are inherently described by its transfer function. For instance, the transfer function of an ARMA($p$, $q$) model with $n$-dimensional parameters $\boldsymbol{\xi}=(\sigma, \lambda_1,\cdots,\lambda_p, \mu_1, \cdots, \mu_q)$ takes the following form:
	\begin{align}
		\label{tf_arma}
		h(z;\boldsymbol{\xi})=\frac{\sigma^2}{2\pi}\frac{(1-\mu_{1}z^{-1})(1-\mu_{2}z^{-1})\cdots(1-\mu_{q}z^{-1})}{(1-\lambda_{1}z^{-1})(1-\lambda_{2}z^{-1})\cdots(1-\lambda_{p}z^{-1})},
	\end{align}
	where $\sigma$ is the gain, $\lambda_i$ is a pole, and $\mu_i$ is a zero of the transfer function. The poles and zeros of stationary filters exist inside the unit disk.
	
	From the transfer function $h(z;\boldsymbol{\xi})$, the spectral density function $S(z;\boldsymbol{\xi})$, describing how the energy is distributed by the system, is defined as 
	\begin{align}
		S(z;\boldsymbol{\xi})=|h(z;\boldsymbol{\xi})|^2,
	\end{align}
	where $|\cdot|$ is the modulus of a function.
  
	The logarithmic transfer function is expressed with the complex cepstrum coefficients of a linear system \cite{oppenheim1965superposition}: 
	\begin{align}
		\label{log_transfer_cepstrum}
		\log{h(z;\boldsymbol{\xi})}=\sum_{s=0}^{\infty} c_s(\boldsymbol{\xi}) z^{-s},
	\end{align}
	where $c_s$ is the $s$-th complex cepstrum coefficient of the transfer function. Since $c_s$ is the coefficient regarding the basis of $z^{-s}$, it is obtained from
	\begin{align}
		\label{cepstrum_coeff}
		c_s(\boldsymbol{\xi})=\langle \log h(z;\boldsymbol{\xi}), z^{-s} \rangle = \frac{1}{2\pi i}\oint_{\mathbb{T}} \log{h(z;\boldsymbol{\xi})} z^{s}\frac{dz}{z},
	\end{align}
	where $\mathbb{T}$ is the unit circle. 
	
	By plugging Eq. (\ref{tf_arma}) into Eq. (\ref{cepstrum_coeff}), the cepstrum coefficients of the ARMA model are obtained as
	\begin{align}
	\label{cepstrum_coef_arma}
	c_s=\left\{ 
	\begin{array}{ll}
	\frac{\sum_{i=1}^{p+q} \gamma_i (\xi^{(i)})^s}{s} & (s \neq 0)\\ 
	1 & (s=0),
	\end{array}
	\right.
	\end{align}
	where $\gamma_i=-1$ when $\xi^{(i)}$ is a pole and $\gamma_i=1$ for zeros.
	
	 The information geometry of linear systems is found under the conditions on stability, minimum phase, and finite Hardy norm of the logarithmic spectral density function \cite{amari1987differential}. In particular, the last condition is given as follows:
	\begin{align}
		\label{hardy_lsdf}
		\frac{1}{2\pi i}\oint_{\mathbb{T}} |\log{S(z;\boldsymbol{\xi})}|^2 \frac{dz}{z}<\infty,
	\end{align}
	where $\mathbb{T}$ is the unit circle. Eq. (\ref{hardy_lsdf}) indicates that the logarithmic spectral density function is a function in Hardy spaces.
	
	Under the conditions on linear systems mentioned above, the Fisher information matrix and $\alpha$-connection of the linear system information geometry \cite{amari1987differential} are represented with
	\begin{align}
		g_{ij}(\boldsymbol{\xi })&=\frac{1}{2\pi i}\oint_{\mathbb{T}}( \partial_{i}\log {S}) ( \partial_{j}\log {S}) \frac{dz}{z},\\
		\Gamma _{ij,k}^{(\alpha )}(\boldsymbol{\xi })&=\frac{1}{2\pi i}\oint_{\mathbb{T}}(\partial_{i}\partial_{j}\log {S}-\alpha ( \partial_{i}\log{S}) ( \partial _{j}\log {S}) )(\partial_{k}\log {S})\frac{dz}{z},
	\end{align}
	where $\mathbb{T}$ is the unit circle, and $i, j,$ and $k$ runs from 1 to $n$. When $\alpha=0$, the connection is the Levi--Civita connection.
	
	Similar to the spectral density function, the K\"ahler information geometry of linear systems can be constructed in terms of transfer functions \cite{choi2015kahlerian}. Using the transfer function, the finite norm condition corresponding to Eq. (\ref{hardy_lsdf}) is expressed with 
	\begin{align}
		\label{hardy_ltf}
		\frac{1}{2\pi i}\oint_{\mathbb{T}} |\log{h(z;\boldsymbol{\xi})}|^2 \frac{dz}{z}<\infty,
	\end{align}
	where $\mathbb{T}$ is the unit circle. Eq. (\ref{hardy_ltf}) indicates that the logarithmic transfer function is a function in Hardy spaces.
	
	According to Choi and Mullhaupt \cite{choi2015kahlerian}, the linear system geometry from transfer functions is the K\"ahler manifold of which the Fisher information matrix and Levi--Civita connection are given as
	\begin{align}
		\label{metric_tf}
		g_{i\bar{j}}(\boldsymbol{\xi},\bar{\boldsymbol{\xi}})&=\frac{1}{2\pi i}\oint_{\mathbb{T}}\big( \partial_i\log{h(z;\boldsymbol{\xi})}\big)\overline{\big( \partial_{j}\log{h(z;\boldsymbol{\xi})\big)}}\frac{dz}{z},\\
		\label{connection_tf}
		\Gamma^{LC}_{ij,\bar{k}}(\boldsymbol{\xi},\bar{\boldsymbol{\xi}})&=\frac{1}{2\pi i}\oint_{\mathbb{T}}\big(\partial_i\partial_j \log{h(z;\boldsymbol{\xi})}\big)\overline{\big(\partial_k\log{h(z;\boldsymbol{\xi})}\big)}\frac{dz}{z},
	\end{align}
	where $\mathbb{T}$ is the unit circle, $i, j,$ and $k$ runs from 1 to $n$, an unbarred index is from holomorphic coordinates, and a barred index is for anti-holomorphic coordinates. $\alpha$-connection is also given in Choi and Mullhaupt \cite{choi2015kahlerian}. For the definition of holomorphic/anti-holomorphic coordinates and more details on K\"ahler manifolds, please refer to the related literature \cite{choi2015application,choi2015kahlerian}.
	
	In K\"ahler geometry, the non-vanishing components of the metric tensor and the Levi--Civita connection are obtained from K\"ahler potential via the following forms:
	\begin{align}
		\label{metric_hardy}
		g_{i\bar{j}}(\boldsymbol{\xi},\bar{\boldsymbol{\xi}})&=\partial_i \partial_{\bar{j}}\mathcal{K},\\
		\label{connection_hardy}
		\Gamma_{ij,\bar{k}}^{LC}(\boldsymbol{\xi},\bar{\boldsymbol{\xi}})&=\partial_i\partial_{j} \partial_{\bar{k}}\mathcal{K},
	\end{align}
	where $\mathcal{K}$ is the K\"ahler potential of the manifold.
	
	It is noted that the K\"ahler potential for the K\"ahler information geometry of linear systems corresponds to the squared Hardy norm of logarithmic transfer functions \cite{choi2015kahlerian}:
	\begin{align}
		\label{khaler_potential_hardy}
		\mathcal{K}=\|\log{h(z;\boldsymbol{\xi})}\|_{H^2}^2=\frac{1}{2\pi i}\oint_{\mathbb{T}} |\log{h(z;\boldsymbol{\xi})}|^2 \frac{dz}{z},
	\end{align}
	where $\mathbb{T}$ is the unit circle. It is easy to derive Eq. (\ref{khaler_potential_hardy}) by comparing Eq. (\ref{metric_tf}) and Eq. (\ref{connection_tf}) with Eq. (\ref{metric_hardy}) and Eq. (\ref{connection_hardy}), respectively. By plugging Eq. (\ref{log_transfer_cepstrum}) to Eq. (\ref{khaler_potential_hardy}), the K\"ahler potential also corresponds to the squared cepstrum norm expressed with cepstrum coefficients $c_s$:
	\begin{align}
    \label{kahler_potential_unweighted}
		\mathcal{K}=\sum_{s=0}^{\infty} |c_s|^2.
	\end{align}
    Eq. (\ref{kahler_potential_unweighted}) means that the K\"ahler potential is the squared cepstrum norm of the system.
    
    Let us consider the extension of the discussion given above to weighted Hardy spaces as done in Choi \cite{choi2021k}. For a weight vector $\omega=(\omega_0, \omega_1,\cdots)$ such that $\omega_i$ is nonnegative, the weighted Hardy norm ($H^2_\omega$-norm) \cite{paulsen2016introduction} is defined as
	\begin{align}
		\|f\|_{H^2_\omega}=\|f\|_{\omega}=\bigg(\sum_{s=0}^{\infty} \omega_s |f_s|^2\bigg)^{1/2},
	\end{align}
	where $f_s$ is the $s$-th Fourier (or Z-transformed) coefficient of $f$, and $\omega_s$ is the $s$-th weight sequence of the weight vector $\omega$. 
	
	The relation between well-known function spaces and weighted Hardy spaces was presented in Choi \cite{choi2021k} as follows:
	\begin{align}
		\label{various_weighted_hardy_norms}
		\omega_s=\left\{ 
		\begin{array}{ll}
			1 & \textrm{for unweighted Hardy space } H^2\\ 
			1+s^2+s^4+\cdots+s^{2m} & \textrm{for Sobolev space } \mathcal{W}^{m,2}\\
			s & \textrm{for Dirichlet space } \mathcal{D}\\
			\frac{1}{1+s} & \textrm{for Bergman space } \mathcal{A}\\
			s^{m} & \textrm{for differentiation semi-norm space } \tilde{\mathcal{D}}^m
		\end{array}
	\right.
	\end{align}
	for a nonnegative integer $s$.
	
	The weighted Hardy norm can be applied to a smooth transformation $\phi$ of a transfer function $h$ \cite{choi2021k}:
	\begin{align}
        \label{weighted_hardy_general_norm}
		\mathcal{I}_{\omega}=\| \phi\circ h \|_{\omega}=\bigg(\sum_{s=0}^{\infty} \omega_s | f_s |^2\bigg)^{1/2},
	\end{align}
	where $f_s$ is the $s$-th Fourier coefficient of $f=\phi\circ h$. 
	
	In the same setting, the distance between two linear systems based on the weighted Hardy norm \cite{choi2021k} is defined as
	\begin{align}
		\label{weighted_hardy_general_dist}
		\mathcal{I}_{\omega}(M, M')=\| \phi\circ h-\phi\circ h' \|_{\omega}=\bigg(\sum_{s=0}^{\infty} \omega_s | f_{s}-f_{s}' |^2\bigg)^{1/2},
	\end{align}
	where $f_{s}$ and $f_{s}'$ are the $s$-th Fourier coefficients of $\phi\circ h$ and $\phi\circ h'$, respectively. When $f_s'=0$ for all $s$, Eq. (\ref{weighted_hardy_general_dist}) is reduced to Eq. (\ref{weighted_hardy_general_norm}).
	
	As explained in Choi \cite{choi2021k}, a logarithmic function is a popular and meaningful candidate for $\phi$. When $\phi(x)=\log{x}$, the weighted Hardy norm of a logarithmic transfer function is represented with
	\begin{align}
		\label{weighted_cepstrum_norm}
		\mathcal{I}_\omega=\bigg(\sum_{s=0}^{\infty} \omega_s|c_s|^2\bigg)^{1/2},
	\end{align}
	where $c_{s}$ is the $s$-th cepstrum coefficient of the transfer function $h$.
	
	Similarly, the weighted Hardy distance between two linear systems (or the weighted cepstrum norm) \cite{choi2021k} is expressed with
	\begin{align}
		\label{weighted_cepstrum_distance}
		\mathcal{I}_{\omega}(M, M')=\| \log h-\log h' \|_{\omega}=\bigg(\sum_{s=0}^{\infty} \omega_s | c_{s}-c_{s}' |^2\bigg)^{1/2},
	\end{align}
	where $c_{s}$ and $c_{s}'$ are the $s$-th cepstrum coefficient of $h$ and $h'$, respectively. It is noteworthy that the weighted Hardy norm of a logarithmic transfer function given by Eq. (\ref{weighted_cepstrum_norm}) is the distance between the linear system and the identity filter, Eq. (\ref{weighted_cepstrum_distance}) where $h'=1$ \cite{choi2021k}. 
	
	According to Choi \cite{choi2021k}, the induced geometry of Eq. (\ref{weighted_cepstrum_distance}) is K\"ahler geometry and derived from K\"ahler potential. The non-vanishing components of the metric tensor and the Levi--Civita connection are given in the following forms:
	\begin{align}
		\label{metric_weighted_hardy}
		g_{i\bar{j}}(\boldsymbol{\xi},\bar{\boldsymbol{\xi}};\omega)&=\partial_i \partial_{\bar{j}}\mathcal{K}_{\omega},\\
		\label{connection_weighted_hardy}
		\Gamma_{ij,\bar{k}}^{LC}(\boldsymbol{\xi},\bar{\boldsymbol{\xi}};\omega)&=\partial_i\partial_{j} \partial_{\bar{k}}\mathcal{K}_{\omega},
	\end{align}
	where $\mathcal{K}_{\omega}$ is the K\"ahler potential of the manifold in the form of
	\begin{align}
    \label{kahler_potential_weighted}
		\mathcal{K}_{\omega}=\|\log{h(z;\boldsymbol{\xi})}\|_{\omega}^2=\sum_{s=0}^{\infty} \omega_s |c_s|^2.
	\end{align}
	From Eq. (\ref{kahler_potential_weighted}), the square of the weighted Hardy norm corresponds to the K\"ahler potential of the linear system geometry in weighted Hardy spaces. It is also considered as a weighted cepstrum norm, the generalized version of Eq. (\ref{khaler_potential_hardy}) and Eq. (\ref{kahler_potential_unweighted}). When the unit weight sequence is used, the K\"ahler potential, Eq. (\ref{kahler_potential_weighted}), is reduced to the K\"ahler potential for the unweighted Hardy space, (\ref{khaler_potential_hardy}) and Eq. (\ref{kahler_potential_unweighted}).
	
	In case of $\omega_s=s$, the weighted cepstrum norm of a linear system is Dirichlet norm \cite{choi2021k}. The Dirichlet norm of the logarithmic transfer function of a linear system is represented by
	\begin{align}
		\mathcal{I}=\| \log{h} \|_{\mathcal{D}}=\bigg(\sum_{s=1}^{\infty} s|c_s|^2\bigg)^{1/2},
	\end{align}
	where $c_{s}$ is the $s$-th cepstrum coefficient of a transfer function $h$. For an ARMA model, this norm is also known as the mutual information between past and future \cite{martin2000metric}. 
	
	The Dirichlet distance between two linear systems, $M$ and $M'$, is given as
	\begin{align}
		\label{dirichlet_dist}
		\mathcal{I}(M, M')=\| \log h-\log h' \|_{\mathcal{D}}=\bigg(\sum_{s=1}^{\infty} s | c_{s}-c_{s}' |^2\bigg)^{1/2},
	\end{align}
	where $c_{s}$ and $c_{s}'$ are the $s$-th cepstrum coefficient of $h$ and $h'$, respectively.
	
	The K\"ahler information manifold induced from Dirichlet norm and distance is also found by Choi \cite{choi2021k}. The metric tensor and connection of the induced geometry are also derived from Eq. (\ref{kahler_potential_weighted}) by using Eq. (\ref{metric_weighted_hardy}) and Eq. (\ref{connection_weighted_hardy}).
	
\section{Hyperbolic decomposition of Dirichlet distance between two ARMA models}
\label{geometric_interpretation}
	In this section, we present the hyperbolic decomposition of Dirichlet distance for ARMA processes which can be also reduced to AR processes and MA processes.

	According to Choi \cite{choi2021k}, the weighted cepstrum distance $\mathcal{I}(M, M')$ between two ARMA($p,q$) and ARMA($p',q'$) models, $M$ and $M'$, in Dirichlet space is expressed with 
	\begin{align}
	\label{dirichlet_distance_arma}
		\mathcal{I}^2(M, M')&=\sum_{s=1}^{\infty} \frac{|(\sum_{i=1}^{p}\lambda _{i}^{s}-\sum_{i=1}^{q}\mu _{i}^{s})-(\sum_{i=1}^{p'}\lambda_{i}'^{s}-\sum_{i=1}^{q'}\mu_{i}'^{s})|^2}{s},
	\end{align}
	where $\lambda_i$ ($\mu_i$) and $\lambda_i'$ ($\mu_i'$) are poles (roots) of the ARMA($p,q$) and ARMA($p',q'$) models, respectively. This Dirichlet distance $\mathcal{I}(M, M')$ can be expressed with logarithmic functions as
	\begin{align}
    \label{dirichlet_distance_arma_log}
    \begin{aligned}
		\mathcal{I}^2(M, M')=&\log{\bigg(\frac{\prod_{i=1}^{p}\prod_{j=1}^{p'}|1-\lambda_i\bar{\lambda}_j'|^2}{\prod_{i=1}^{p}\prod_{j=1}^{p}(1-\lambda_i\bar{\lambda}_j)\prod_{i=1}^{p'}\prod_{j=1}^{p'}(1-\lambda_i'\bar{\lambda}_j')}\bigg)}\\
		&+\log{\bigg(\frac{\prod_{i=1}^{q}\prod_{j=1}^{q'}|1-\mu_i\bar{\mu}_j'|^2}{\prod_{i=1}^{q}\prod_{j=1}^{q}(1-\mu_i\bar{\mu}_j)\prod_{i=1}^{q'}\prod_{j=1}^{q'}(1-\mu_i'\bar{\mu}_j')}\bigg)}\\
		&+\log{\bigg(\frac{\prod_{i=1}^{p}\prod_{j=1}^{q}|1-\lambda_i\bar{\mu}_j|^2\prod_{i=1}^{p'}\prod_{j=1}^{q'}|1-\lambda_i'\bar{\mu}_j'|^2}{\prod_{i=1}^{p}\prod_{j=1}^{q'}|1-\lambda_i\bar{\mu}_j'|^2\prod_{i=1}^{p'}\prod_{j=1}^{q}|1-\lambda_i'\bar{\mu}_j|^2}\bigg)}.
    \end{aligned}
    \end{align}
    The equation given above can be decomposed into the following terms:
    \begin{align}
    \label{dirichlet_distance_arma_decomp}
		\mathcal{I}^2(M, M')=\mathcal{I}^2_{AR-AR}+\mathcal{I}^2_{MA-MA}+\mathcal{I}^2_{AR-MA},
    \end{align}
    where $\mathcal{I}^2_{AR-AR}$ is the terms from the AR parts from two ARMA models, $\mathcal{I}^2_{MA-MA}$ is constructed from the MA parts from the ARMA models, and $\mathcal{I}^2_{AR-MA}$ is the cross terms between the AR parts and the MA parts.

    It is important to note that we can obtain an expression similar to Eq. (\ref{dirichlet_distance_arma}) and Eq. (\ref{dirichlet_distance_arma_log}) for AR or MA models by setting no zeros or no poles, respectively. It is clear to check from Eq. (\ref{dirichlet_distance_arma_decomp}). Only the AR (or MA) parts of the distance are considered for AR (or MA) models. Additionally, we can also derive the Dirichlet norm of the ARMA model from Eq. (\ref{dirichlet_distance_arma_log}) by excluding poles and zeros from one of the models. The Dirichlet norm of the AR (or MA) model can be obtained in similar ways.
  
	With the coordinate system of $\boldsymbol{\xi}=(\xi^{(1)}=\lambda_1, \cdots,\xi^{(p)}=\lambda_p, \xi^{(p+1)}=\mu_1,\cdots,\xi^{(p+q)}=\mu_q)$, the metric tensor and the connection components of the K\"ahler-ARMA geometry are also given by Choi \cite{choi2021k} as 
	\begin{align}
		g_{i\bar{j}}&=\gamma_i\gamma_j\frac{1}{(1-\xi^{(i)}\bar{\xi}^{(j)})^2},\\
		\Gamma_{ij,\bar{k}}&=\gamma_j\gamma_k \frac{2\bar{\xi}^{(k)}}{(1-\xi^{(j)}\bar{\xi}^{(k)})^3}\delta_{ij},
	\end{align}
	where $i,j,k$ runs from $1$ to $p+q$, $\delta_{ij}$ is the Kronecker delta, and $\gamma_i$ is $-1$ for poles and 1 for zeros.

	It is noteworthy that the Dirichlet distance between two ARMA($p,q$) and ARMA($p',q'$) models, Eq. (\ref{dirichlet_distance_arma}) and Eq. (\ref{dirichlet_distance_arma_log}), is expressed with functions in hyperbolic distance. The hyperbolic distance between $u$ and $v$ inside the hyperbolic disk is defined as
	\begin{align}
		\rho(u,v)=\log{\frac{|1-u\bar{v}|+|u-v|}{|1-u\bar{v}|-|u-v|}}.
	\end{align} 
	
	From the hyperbolic distance $\rho$, we define $\Xi$ as follows:
	\begin{align}
	\label{xi}
		\Xi(u,v)=\log{\big(\cosh^2{[\frac{1}{2}\rho(u,v)]}\big)}=\log{\big(\frac{1}{2}(\cosh{\rho(u,v)}+1)\big)},
	\end{align}
	and $\Xi$ can be expressed directly with $u$ and $v$: 
	\begin{align}
		\Xi(u,v)=\log{|1-u\bar{v}|^2} - \log{(1-|u|^2)} -\log{(1-|v|^2)}.
	\end{align}
	In contrast to $\rho$, $\Xi(u,v)$ is not a distance measure. $\Xi(u,v)$ is non-negative because $\cosh^2{[\frac{1}{2}\rho(u,v)]}\ge1$ for all $u$ and $v$ inside the hyperbolic disk. Moreover, it is symmetric under $u$ and $v$. However, the triangle inequality is not satisfied. 
	
	The Dirichlet distance $\mathcal{I}(M, M')$ between two ARMA models can be written in terms of Eq. (\ref{xi}):
	\begin{align}
    \label{dirichlet_dist_arma}
    \begin{aligned}
		&\mathcal{I}^2(M, M')\\
        =&\sum_{i,j=1}^{p,p'}\Xi(\lambda_i,\lambda_j')-\sum_{i<j}^{p}\Xi(\lambda_i,\lambda_j)-\sum_{i<j}^{p'}\Xi(\lambda_i',\lambda_j')\\
		&+\sum_{i,j=1}^{q,q'}\Xi(\mu_i,\mu_j')-\sum_{i<j}^{q}\Xi(\mu_i,\mu_j)-\sum_{i<j}^{q'}\Xi(\mu_i',\mu_j')\\
		&+\sum_{i,j=1}^{p,q}\Xi(\lambda_i,\mu_j)+\sum_{i,j=1}^{p',q'}\Xi(\lambda_i',\mu_j')-\sum_{i,j=1}^{p,q'}\Xi(\lambda_i,\mu_j')-\sum_{i,j=1}^{q,p'}\Xi(\mu_i,\lambda_j')\\
		&+\big((p-q)-(p'-q')\big)\log{\bigg(\frac{\prod_{i=1}^{p'} (1-|\lambda_i'|^2)\prod_{i=1}^q (1-|\mu_i|^2)}{\prod_{i=1}^p (1-|\lambda_i|^2)\prod_{i=1}^{q'} (1-|\mu_i'|^2)}\bigg)}.
	\end{aligned}
    \end{align}
	The first three terms come from the AR parts, and the next three terms are from the MA parts. The next four terms are related to the mixing of AR and MA parts. It is noteworthy that if we use the same relative orders between AR parts and MA parts across two models, the last term in Eq. (\ref{dirichlet_dist_arma}) vanishes and the Dirichlet distance is completely expressed with the functions depending on the hyperbolic distances. We can also find the hyperbolic decomposition for the Dirichlet norm in similar ways.
	
	Since AR models and MA models are submodels of ARMA models, it is straightforward to derive the Dirichlet distances and hyperbolic decomposition for AR models and MA models from ARMA models. It is also possible to obtain the hyperbolic decomposition for the Dirichlet norm.

    This decomposition can be used for model order reduction. Let us assume that we have an ARMA model with known $p'$ poles and $q'$ zeros, and we try to find a smaller ARMA ($p$, $q$) models, i.e., $p<p'$ and $q<q'$, but close to the original ARMA model. If their relative orders between AR poles and MA zeros are identical, i.e., $p-q=p'-q'$, the last term in Eq. (\ref{dirichlet_dist_arma}) becomes zero. Moreover, since the terms, related to purely $p'$ poles and $q'$ zeros in Eq. (\ref{dirichlet_dist_arma}), are known, the number of terms we need to consider is significantly reduced to seven from eleven, and this is an advantage to minimize the Dirichlet distance with the poles and zeros from the smaller ARMA model.
    
\section{Conclusion}
	We find that the Dirichlet distance, the weighted Hardy distance with $\omega_s=s$, between two ARMA models is decomposed into functions of the hyperbolic distances between the poles and roots of an ARMA transfer function. Additionally, the Dirichlet distance can be expressed with pure AR terms, pure MA terms, and AR-MA mixing terms between two models. The decomposition is useful for model order reduction to find ARMA models with fewer parameters to preserve the behavior of ARMA models with more parameters.
 
\bibliographystyle{plain}
\bibliography{HyperbolicDirichletARMA}

\begin{thebibliography}{10}

\bibitem{amari1987differential}
Shun-ichi Amari.
\newblock Differential geometry of a parametric family of invertible linear systems—{R}iemannian metric, dual affine connections, and divergence.
\newblock {\em Mathematical systems theory}, 20(1):53--82, 1987.

\bibitem{barbaresco2006information}
Fr{\'e}d{\'e}ric Barbaresco.
\newblock Information intrinsic geometric flows.
\newblock In {\em AIP Conference Proceedings}, volume 872, pages 211--218. American Institute of Physics, 2006.

\bibitem{choi2021k}
Jaehyung Choi.
\newblock K{\"a}hler information manifolds of signal processing filters in weighted hardy spaces.
\newblock {\em arXiv preprint arXiv:2108.07746}, 2021.

\bibitem{choi2015application}
Jaehyung Choi and Andrew~P Mullhaupt.
\newblock Application of {K}{\"a}hler manifold to signal processing and {B}ayesian inference.
\newblock In {\em AIP Conference Proceedings}, volume 1641, pages 113--120. American Institute of Physics, 2015.

\bibitem{choi2015geometric}
Jaehyung Choi and Andrew~P Mullhaupt.
\newblock Geometric shrinkage priors for {K}{\"a}hlerian signal filters.
\newblock {\em Entropy}, 17(3):1347--1357, 2015.

\bibitem{choi2015kahlerian}
Jaehyung Choi and Andrew~P Mullhaupt.
\newblock K{\"a}hlerian information geometry for signal processing.
\newblock {\em Entropy}, 17(4):1581--1605, 2015.

\bibitem{martin2000metric}
Richard~J Martin.
\newblock A metric for {ARMA} processes.
\newblock {\em IEEE transactions on Signal Processing}, 48(4):1164--1170, 2000.

\bibitem{oda2021shrinkage}
Hidemasa Oda and Fumiyasu Komaki.
\newblock Shrinkage priors on complex-valued circular-symmetric autoregressive processes.
\newblock {\em IEEE Transactions on Information Theory}, 67(8):5318--5333, 2021.

\bibitem{oppenheim1965superposition}
Alan~V Oppenheim.
\newblock Superposition in a class of nonlinear systems.
\newblock 1965.

\bibitem{paulsen2016introduction}
Vern~I Paulsen and Mrinal Raghupathi.
\newblock {\em An introduction to the theory of reproducing kernel {H}ilbert spaces}, volume 152.
\newblock Cambridge university press, 2016.

\bibitem{ravishanker2001differential}
Nalini Ravishanker.
\newblock Differential geometry of {ARFIMA} processes.
\newblock {\em Communications in Statistics-Theory and Methods}, 30(8-9):1889--1902, 2001.

\bibitem{ravishanker1990differential}
Nalini Ravishanker, Edward~L Melnick, and Chih-Ling Tsai.
\newblock Differential geometry of {ARMA} models.
\newblock {\em Journal of Time Series Analysis}, 11(3):259--274, 1990.

\bibitem{tanaka2018superharmonic}
Fuyuhiko Tanaka.
\newblock Superharmonic priors for autoregressive models.
\newblock {\em Information Geometry}, 1(2):215--235, 2018.

\end{thebibliography}

\end{document}